\documentclass[aps,prd,showpacs,nofootinbib,12pt]{revtex4}

\usepackage{amssymb}
\usepackage{amsmath}
\usepackage{bm}
\expandafter\ifx\csname package@font\endcsname\relax\else
\expandafter\expandafter
 \expandafter\usepackage
 \expandafter\expandafter
 \expandafter{\csname package@font\endcsname}%
\fi
\renewcommand{\prd}{{\it Phys. Rev. D}}
\renewcommand{\prl}{{\it Phys. Rev. Lett.}}

\newcommand{\physrep}{{\it Phys. Rep.}}

\begin{document}
\title{The gravitomagnetic influence on Earth-orbiting spacecrafts and on the lunar orbit}
\author{Sergei M. Kopeikin}
\email{kopeikins@missouri.edu}
\affiliation{Department of Physics \& Astronomy, University of
Missouri-Columbia, 65211, USA}
\begin{abstract}
Gravitomagnetic field is covariantly split in the {\it intrinsic} and {\it extrinsic} parts, which are generated by rotational and translational currents of matter respectively. The {\it intrinsic} component has been recently discovered in the LAGEOS spacecraft experiment. We discuss the method of detection of the {\it extrinsic} tidal component with the lunar laser ranging (LLR) technique. Analysis of the gauge residual freedom in the relativistic theory of three-body problem demonstrates that LLR is currently not capable to detect the {\it extrinsic} gravitomagnetic effects which are at the ranging level of few millimeters. Its detection requires further advances in the LLR technique that are coming in the next 5-10 years.
\end{abstract}
\maketitle\noindent
Detection of gravitomagnetic components as predicted by Einstein's theory of general relativity is one of the primary goals of experimental gravitational physics.
The paper \cite{nortur} states that the gravitomagnetic interaction plays a part in shaping the lunar orbit readily obervable by LLR. The authors picked up a ``gravitomagnetic" term from the parameterized post-Newtonian (PPN) equation of motion of massive bodies \cite{will} and proved that it correctly reproduces the Lense-Thirring precession of the GP-B gyroscope. The paper \cite{nortur} argues that the very same term in the equations of motion of the Moon, derived in the solar-system barycentric (SSB) frame, perturbs the lunar orbit with a radial amplitude $\simeq 6$ meters that was observed with the lunar laser ranging (LLR). We have explained \cite{kop-prl} and confirm in more detail in the present paper that the gauge freedom of the relativistic three-body problem suppresses the gravitomagnetic effects in the lunar motion, that depend on the Earth's velocity ${\bm V}$ around the Sun, to the level $\le 1$ centimeter. It makes LLR currently insensitive to the gravitomagnetic interaction.

There are two types of mass currents in gravity \cite{ciu,kop-ijmpd}. The first type is
produced by the intrinsic rotation of matter around body's center of
mass. It generates an {\it intrinsic} gravitomagnetic field tightly
associated with body's angular momentum (spin) and most research in
gravitomagnetism has been focused on the discussion of its various
properties. Textbook \cite{cw} gives a comprehensive review of various aspects of the {\it intrinsic} gravitomagnetism. It is interesting to note that the {\it intrinsic} gravitomagnetic field can be associated with the holonomy invariance group \cite{mmm}. Some authors \cite{cam1,cam2} have proposed to measure the {\it intrinsic} gravitomagnetic field by observing quantum effects of coupling of fermion's spin with the angular momentum of the Earth. It might be worthwhile to explore association of the {\it intrinsic} gravitomagnetism with the classic Hannay precession phase \cite{hannay,spal1,spal2}.

The first classic experiment to test the {\it intrinsic} gravitomagnetic effect of the rotating Earth has been carried out by observing LAGEOS in combination with other geodetic satellites \cite{ciuf,cp,cp-3,cf-nat} which verified its existence with a remarkable precision as predicted by
Einstein's general relativity. Independent experimental measurement of the {\it intrinsic} gravitomagnetic field of the rotating Earth is currently under way by
the Gravity Probe B mission \cite{gpb}.

The second type of the mass
current is caused by translational motion of matter. It generates an
{\it extrinsic} gravitomagnetic field that depends on the frame of
reference of observer and can be either completely eliminated in the rest
frame of the matter or significantly suppressed by the transformation to the local inertial reference frame of observer. This property of the {\it extrinsic} gravitomagnetic
field is a direct consequence of the gauge invariance of Einstein's
gravity field equations for an isolated astronomical system \cite{fock-2} embedded to the asymptotically-flat space-time. Experimental testing of the {\it extrinsic} gravitomagnetic field is as important
as that of the {\it intrinsic} gravitomagnetic field. The point is
that both the {\it intrinsic} and the {\it extrinsic} gravitomagnetic
fields obey the same equations and, therefore, their measurement would
essentially complement each other \cite{kop-ijmpd}. Furthermore, detection of the {\it
extrinsic} gravitomagnetic field probes the time-dependent behavior of the
gravitational field which is determined by the structure of the gravity null cone (the domain of causal influence) on which the gravity force propagates. Experimental verification of the gravitomagnetic properties of gravity is important for the theory of braneworlds \cite{bbv} and for setting other, more stringent limitations on vector-tensor theories of gravity \cite{vtni}.

Ciufolini \cite{ciu} proposed to distinguish the rotationally-induced gravitomagnetic field from the translationally-induced gravitomagnetic effects by making use of two scalar invariants of the curvature tensor
\begin{eqnarray}
\label{krin1}
I_1&=&R_{\alpha\beta\mu\nu}R^{\alpha\beta\mu\nu}\;,\\
\label{krin2}
I_2&=&R_{\alpha\beta}{}^{\mu\nu}R^{\alpha\beta\rho\sigma}E_{\mu\nu\rho\sigma}\;,
\end{eqnarray}
where $R_{\alpha\beta\mu\nu}$ is the curvature tensor, $E_{\mu\nu\rho\sigma}$ is the fully anti-symmetric Levi-Civita tensor with $E_{0123}=+\sqrt{-g}$, and $g={\rm det}(g_{\mu\nu})<0$ is the determinant of the metric tensor.
Ciufolini \cite{ciu} notices that a weak gravitational field of an isolated astronomical system yields $I_2=0$ if the {\it intrinsic} gravitomagnetic field is absent. However, one should not confuse the invariant $I_2$ with the gravitomagnetic field itself. The gravitomagnetic field is generated by {\it any} current of matter. Hence, $I_2=0$ does not mean that any gravitomagnetic field is absent as has been erroneously interpreted in \cite{pask}. Equality $I_2=0$ only implies that the gravitomagnetic field is of the {\it extrinsic} origin ($I_1\not=0$), that is generated by a translational motion of matter. The translational gravitomagnetic field can be measured, for instance, by observing the gravitational deflection of light by a moving massive body like Jupiter \cite{kop-pla,kopmak}. This gravitomagnetic frame-dragging effect on the light ray was indeed observed in a dedicated radio-interferometric experiment \cite{fk-iau}.

Paper \cite{nortur} makes an attempt to demonstrate that the {\it extrinsic} gravitomagnetic field can be measured by making use of the LLR observations of the lunar orbit. This must not be confused with the measurement of the {\it intrinsic} gravitomagentic field by means of the satellite laser ranging technique applied to LAGEOS \cite{ciuf,cp,cp-3,cf-nat}. The LAGEOS experiment measures the Lense-Thirring precession of the satellite's orbit caused by the Earth's angular momentum entering $g_{0i}$ component of the metric tensor. The authors of \cite{nortur} have been trying to measure the gravitomagnetic precession of the lunar orbit caused by the orbital motion of the Earth-Moon system around the Sun. They used the barycentric coordinates of the solar system (BCRS) to derive the equation of motion of the Moon relative to the Earth. The equation is effectively obtained as a difference between the Einstein-Infeld-Hoffmann (EIH) equations of motion for the Earth and for the Moon with respect to the barycenter of the solar system, and repeats the original derivation by Brumberg \cite{br-luna,brum-book1}, which later was independently derived by Baierlein \cite{baier}.

The barycentric equation of motion of the Moon formally includes the gravitomagnetic perturbation in the following form \cite{nortur}
\begin{equation}
\label{1}
{\bm a}_{\rm GM}={\bm a}+\frac{\gamma-1}{2}{\bm a}\;,
\end{equation}
where $\gamma$ parameterizes a deviation from general relativity, the bold letters denote spatial vectors, and the dot between two vectors means their Euclidean dot product. The general-relativistic post-Newtonian acceleration
\begin{equation}
\label{2}
{\bm a}=\frac{4Gm}{c^2r^2}\left[\hat{\bm r}\left({\bm V}\cdot{\bm u}\right)-{\bm V}\left({\bm u}\cdot\hat{\bm r}\right)\right]\;,
\end{equation}
where $m$ is mass of the Earth, $r$ is radius of the lunar orbit, $\hat{\bm r}$ is the unit vector from the Earth to the Moon, ${\bm V}$ is the Earth's velocity around the Sun, and ${\bm u}$ is the Moon's velocity around the Earth.

We notice \cite{kop-prl} that the barycentric coordinate frame referred to the geocenter by a simple, Newtonian-like spatial translation (the time coordinate is unchanged)
\begin{equation}\label{3}
{\bm r}={\bm x}-{\bm x}_E(t)\;,
\end{equation}
as it is obtained in \cite{nortur,wtb}, is {\it not} in a free fall about the Sun, and does {\it not} make a local inertial frame. Thus, perturbations in Eqs. (\ref{1})-(\ref{2}) can not be interpreted as physically observable and, in fact, represent a spurious gauge-dependent effect that is canceled by transformation to the local-inertial frame of the geocenter.
This transformation is a generalized Lorentz boost with taking into account a number of additional terms due to the presence of the external gravitational field of the Sun \cite{kop88,kv}

 The gauge freedom of the lunar equations of motion must be analyzed to eliminate all gauge-dependent, non-observable terms. Only the terms in the equations of motion, which can not be eliminated by the gauge transformation to the local inertial frame can be physically interpreted. The analysis of the gauge freedom in the three body-problem had been done in \cite{bk,dsx,kv}. It proves that all non-tidal and ${\bm V}$-dependent terms, including the first term in the right side of Eq. (\ref{1}), are pure coordinate effects that disappear from the lunar equations of motion after transformation to the geocentric, locally-inertial frame. This is because the Lorentz invariance and the principle of equivalence reduce the relativistic equation of motion of the Moon to the covariant equation of the geodesic deviation between the Moon's and the Earth's world lines \cite{bk,dsx}, where gravitomagnetic effects appear only as tidal relativistic forces with amplitude smaller than 1 centimeter. The covariant nature of gravity tells us \cite{will} that if some effect is not present in the local frame of observer, it can not be observed in any other coordinate system. This means that besides physically-observable terms, the barycentric LLR model \cite{nortur,wtb} also operates with terms having the gauge-dependent origin, which mathematically nullify each other in the data-processing computer code irrespectively of the frame of reference. The mutually annihilating terms enter different parts of the barycentric LLR model with opposite signs \cite{bk,kop-prl} but, if taken separately, can be erroneously interpreted as really observable. This is what exactly happened with the misleading analysis given in \cite{nortur}.

General relativity indicates that the barycentric EIH lunar equations of motion may admit the observable gravitomagnetic acceleration only in the form of the second term in the right side of Eq. (\ref{1}) that is proportional to $\gamma-1$. Radio experiments set a limit on $\gamma-1 \le 10^{-3}$ \cite{will} that yields $|{\bm a}_{\rm GM}|\le 1$ millimeter. The current half-centimeter accuracy of LLR is insufficient to measure such negligible effect. Moreover, our analysis of the gauge invariance of the scalar-tensor theory of gravity \cite{kv} points out that the term being proportional to $\gamma$ in equation (\ref{1}) is also eliminated in the locally-inertial, geocentric reference frame. We conclude that LLR is currently insensitive to the gravitomagnetism and, yet, can not compete with the LAGEOS and/or GP-B experiments.

Recent paper by Soffel et al. \cite{skmb} is another attempt to prove that the {\it extrinsic} gravitomagnetic acceleration (\ref{2}) can be measured with the LLR technique in a locally-inertial reference frame. The authors of \cite{skmb} accept our criticism \cite{kop-prl} but continue to believe that one can measure the {\it extrinsic} gravitomagnetic field by making use of the preferred frame parametrization of the gravimagnetic terms. To this end, Soffel et al. \cite{skmb} introduce the preferred-frame generalization of equation (\ref{1}) by replacing
\begin{equation}
\label{4}
\gamma-1\rightarrow \gamma-1+\eta_G/4\;,
\end{equation}
where $\eta_G$ is a parameter labeling the gravitomagnetic effects in the barycentric equations of motion of the Moon. Parameter $\eta_G=-\alpha_1/2$ in the framework of the PPN formalism \cite{will} but unless this is not stated explicitly one can regard $\eta_G$ as an independent parameter and fit it to LLR data irrespectively of $\alpha_1$. This was done by Soffel et al. \cite{skmb} who had obtained $\eta_G=(0.9\pm 0.7)\times 10^{-3}$.

We argue that this measurement says nothing about the {\it extrinsic} gravitomagnetic field. This is because $\eta_G$ has no any fundamental significance. Its value is not invariant and crucially depends on the choice of the preferred reference frame, which one uses for processing LLR data. In general theory of relativity $\eta_G\equiv 0$ in any frame, that means a vanishing (non-observable) effect. Indeed, analytic calculations given in appendix of the paper \cite{skmb} and in \cite{n1,n2} confirm that in the PPN framework with $\alpha_1= 0$ all gravitomagnetic effects in the motion of the solar system are nullified by the corresponding gravitoelectric effects from other well-established post-Newtonian gravitational potentials. Those calculations reveal that the gravitomagnetic effect described by equation (\ref{1}) is nothing but a symbolic property of the particular coordinate system used for calculations. LLR data fit and the coordinate-dependent limits on $\eta_G$ obtained in paper \cite{skmb} confirm that the gravitational model used in the data processing, is self-consistent. However, it neither means that the {\it extrinsic} gravitomagnetic field was measured nor that general theory of relativity was tested. This is because the gauge-invariance is the main property of a large class of the metric-based gravitational theories and testing the self-consistency of the LLR equations does not single out a specific gravitational theory.

In order to measure the {\it extrinsic} gravitomagnetic field, one has to find the real gravitomagnetic effect in the motion of the Moon, which does not vanish in the framework of general theory of relativity in the locally-inertial frame of observer on the Earth. For this reason, the primary goal of the relativistic theory of the lunar motion is to construct an inertial reference frame along the world-line of the geocenter and to identify the gravitomagnetic effects in this frame. This task was solved in our paper \cite{bk} in the post-Newtonian approximation in the case of a three-body problem (Sun, Earth, Moon) under assumption that the Moon is considered as a test particle. In the quadrupole approximation the metric tensor in the locally-inertial geocentric reference frame $X^\alpha=(cT, {\bm X})$ has the following form
\begin{eqnarray}
\label{5}
G_{00}(T,{\bm X})&=&-1+\frac{2}{c^2}\left[U(T,{\bm X})+Q_pX^p+\frac{3}{2} Q_{pq}X^pX^q\right]+O\left(\frac1{c^4}\right)\;,\\
\label{6}
G_{0i}(T,{\bm X})&=&-\frac{4}{c^3}\left[U^i(T,{\bm X})+\epsilon_{ipk}C_{pq}X^kX^q\right]+O\left(\frac1{c^5}\right)\;,\\
\label{7}
G_{ij}(T,{\bm X})&=&\delta_{ij}+\frac{2}{c^2}\left[U(T,{\bm X})+Q_pX^p+\frac{3}{2} Q_{pq}X^pX^q\right]\delta_{ij}+O\left(\frac1{c^4}\right)\;,
\end{eqnarray}
where $U(T,{\bm X})$ is the Newtonian potential of the Earth, $U^i(T,{\bm X})$ is the post-Newtonian vector-potential of the Earth, $Q_p$ is the acceleration of the geocenter with respect to the geodesic world line, $Q_{pq}$ and $C_{pq}$ are tidal gravitational-force gradients from the Sun, and $\epsilon_{ipk}$ is the fully anti-symmetric symbol.

We notice that $U^i(T,{\bm X})$ represents the {\it intrinsic} gravitomagnetic field of the Earth, which has been measured in the LAGEOS experiment \cite{cp}. It can be shown \cite{ci} that gravitomagnetic vector potential $U^i(T,{\bm X})$ produces a negligibly small acceleration on the lunar orbit, and can be discarded. The tensor potential $C_{pq}$ has the {\it extrinsic} gravitomagnetic origin as it is generated in the geocentric frame by the motion of an external mass (the Sun) with respect to the Earth. This potential is expressed in terms of the orbital velocity of the Earth, $V^i$, and the Newtonian tidal matrix $Q_{pq}$ \cite{bk}
\begin{equation}
\label{8}
C_{pq}=\epsilon_{ikp} \left(V_iQ_{kq}-V_kQ_{iq}+\frac12\delta_{kq}V^jQ_{ij}-\frac12\delta_{iq}V^jQ_{kj}\right)\;,
\end{equation}
where
\begin{equation}
\label{9}
Q_{pq}=\frac{GM}{R^5}\left(3X_pX_q-\delta_{pq}X^2\right)\;,
\end{equation}
$M$ is mass of the Sun, and $R$ is the distance between the Sun and Earth.

The gravitomagnetic tidal force causes a non-vanishing gravitomagnetic acceleration of the Moon, which reads in the locally-inertial geocentric frame as follows (see equation 7.12 from \cite{bk})
\begin{equation}
\label{10}
A^i_{GM}=12\epsilon_{ijk}C_{jq}u^qr^k\;,
\end{equation}
where $u^i$ is the geocentric velocity of the Moon, and $r^i$ is the Earth-Moon radius-vector. This gravitomagnetic acceleration causes a radial oscillations of the lunar orbit that can be estimated by making use of equations (8) and (9). Noticing that the term $Q_{ij}r^j$ is proportional to the Newtonian tidal force from the Sun, which produces the {\it variation} inequality of the lunar orbit \cite{koval}, one obtains
\begin{equation}
\label{11}
|{\bm A}_{GM}|\simeq 10\times {\rm (variation)}\times \frac{Vu}{c^2}\;.
\end{equation}
The variation amounts to $\simeq 3000$ km in radial oscillation \cite{koval}. Hence, the amplitude of the oscillation of the lunar orbit caused by the gravitomagnetic tidal force has an amplitude of about 1 cm. This is unmeasurable with the current LLR data but can be measured in the next 5-10 years with the advent of a millimeter-range LLR technology \cite{apollo}.

\end{document}